\documentclass[12pt,onecolumn,aps,superscriptaddress,
               amsmath,amssymb,nofootinbib]{revtex4}      
\usepackage{graphicx}

\usepackage{graphicx,epsfig}

\usepackage{dcolumn}
\usepackage{bm}

\newcommand{\Z}{{\mathbb Z}}

\begin{document}

\begin{flushright}
\baselineskip=12pt \normalsize
{ACT-03-09},
{MIFP-09-07}\\
\smallskip
\end{flushright}

\title{MSSM-like AdS Flux Vacua with Frozen Open-string Moduli}

\author{Ching-Ming Chen}
\affiliation{George P. and Cynthia W. Mitchell Institute for
Fundamental Physics, Texas A\&M
University,\\ College Station, TX 77843, USA}
\author{Tianjun Li}
\affiliation{George P. and Cynthia W. Mitchell Institute for
Fundamental Physics, Texas A\&M
University,\\ College Station, TX 77843, USA}
\affiliation{Key Laboratory of Frontiers in Theoretical Physics,
  Institute of Theoretical Physics, Chinese Academy of Sciences,
  Beijing 100190, P. R. China
}
\author{Van Eric Mayes}
\affiliation{George P. and Cynthia W. Mitchell Institute for
Fundamental Physics, Texas A\&M
University,\\ College Station, TX 77843, USA}
\author{D.V. Nanopoulos}
\affiliation{George P. and Cynthia W. Mitchell Institute for
Fundamental Physics, Texas A\&M
University,\\ College Station, TX 77843, USA}
\affiliation{Astroparticle Physics Group, Houston
Advanced Research Center (HARC),
Mitchell Campus,
Woodlands, TX~77381, USA; \\
Academy of Athens,
Division of Natural Sciences, 28~Panepistimiou Avenue, Athens 10679,
Greece}

\begin{abstract} 
We construct supersymmetric Pati-Salam  
flux vacua in AdS from intersecting
D6-branes on $T^6/(\Z_2 \times \Z_2')$.  The models
constructed have three generations 
of MSSM matter plus right-handed neutrinos. 
Because the cycles wrapped by the D-branes are 
rigid there are no extra massless fields in the 
adjoint representation, arising as open-string moduli.
However, we find that it is problematic to break the Pati-Salam
gauge symmetry to the Standard Model (SM) while keeping the 
SM hypercharge massless.  
           
\end{abstract}
\maketitle
\newpage
\section{Introduction}

With the dawn of the 
Large Hadron Collider (LHC)
era, the prospects for the discovery of new physics 
may finally be arriving. 
In particular, whatever physics is responsible for stabilizing 
the electroweak scale should be discovered.  Signals of
the favored mechanism, broken supersymmetry, may be observed as 
well as the Higgs states required to break the electroweak symmetry. 
Together with the new windows
opening up with advances in cosmology, 
we may soon be on the road to a detailed
understanding of the universe. 

In principle, it should be possible to 
derive all known physics in a top-down 
approach directly from 
string theory, as well as potentially 
predicting new and unexpected phenomena. 
However, at present there seems to be a tremendous
degeneracy of possible vacua, and in the absence
of any non-perturbative formulation of string theory,
there seems to be no reason for singling out one vacuum
or another.  Conversely, following a bottom-up approach, one may ask
if it is possible to deduce the origin of new
physics given such a signal at LHC.  
For example, in the case of low-energy supersymmetry,
it may be possible from the experimental data to deduce 
the structure of the fundamental theory at high energy scales
which determines the
supersymmetry breaking soft terms and ultimately leads
to electroweak symmetry breaking (EWSB). Although still in initial
stages, there has been some effort put towards this approach
in recent years studying the pattern of signatures of 
broad classes of string vacua~\cite{ArkaniHamed:2005px,Kane:2006yi, Kane:2007pp, Feldman:2007zn, Feldman:2008hs, Maxin:2008kp}.    

Both top-down and bottom-up approaches have various strengths
and weaknesses.  However, it seems likely that progress
will be made by a combination of the two approaches. 
Specifically, by constructing vacua which increasingly
resemble our world, including the hoped for new data gleaned from LHC in
the coming years, it should be possible to at least
restrict the landscape of possible vacua to isolated regions.
Then, from the low-energy effective action of these models it should
be possible to extract experimental signatures 
which may be matched to those observed at LHC.  Thus, for this program
to succeed it is imperative that 
we have concrete, realistic models in hand.  However,
as is well known, at present there is not even one string-derived 
model which can be considered fully realistic. 
Thus, it is difficult to draw any inferences
based on the statistical study of the landscape
since there are no truly representative samples
that match the universe we live in. 

An important question is what criteria must 
a particular model satisfy in order
to be considered fully realistic.  Besides 
obtaining three-generations of
chiral fermions which transform as bifundamental
representations of
$SU(3)_C \times SU(2)_L \times U(1)_Y$, 
such a model should be capable
of reproducing the observed hierarchy of 
fermion masses, including the observed 
small neutrino
masses.  The values of the gauge coupling 
constants should also be predicted.  
Although it has become somewhat fashionable
these days to
dismiss the apparent unification of the gauge 
couplings in the MSSM as accidental,
we still believe that this is an important and 
significant feature that
should be present in any fully realistic model.
Furthermore, it should be possible
to address the issue of supersymmetry breaking 
within the model and extract detailed
predictions for the superpartner spectrum. It 
should also be possible to calculate the $\mu$ and
$B\mu$ terms in the effective action. Combined with a calculation
of the Yukawa couplings for quarks and leptons,
this then allows important parameters such as tan $\beta$
to be determined.  
Mechanisms within the model for realizing 
inflation and dark energy would also be desirable.  
Finally, the importance of
moduli stablization cannot be overstressed as 
all physical parameters such 
as Yukawa and gauge couplings depend crucially 
upon the values taken by the moduli VEVs.

Much work towards model 
building has been focused on the heterotic string. 
Indeed, many of the 
phenomenologically most appealing of such 
models were those constructed within the 
free-fermionic formulation based on the NAHE set~\cite{AEHN}. 
In addition, progress has also been made in recent years in
constructing heterotic M-theory models~\cite{Braun:2005nv, Bouchard:2005ag}.    
On the Type II side,
D-branes~\cite{JPEW} have created new approaches to constructing 
semi-realistic vacua.
In particular, intersecting D-brane worlds, where chiral fermions
are localized at the intersections between D-branes (in the Type IIA
picture) \cite{Berkooz:1996km} 
or in the T-dual (Type IIB)
picture involving magnetized D-branes \cite{Bachas:1995ik}, 
have provided an exciting new avenue in model building.  
Much work has been done in recent years in constructing
such models.  For recent reviews of this topic, see~\cite{Blumenhagen:2005mu} 
and~\cite{Blumenhagen:2006ci}.

Although much progress has been made in constructing
intersecting D-brane worlds, 
none of the models to date have 
been completely satisfactory.  Problems 
include extra chiral and 
non-chiral matter and the lack of a complete set 
of Yukawa couplings, 
which are typically forbidden by global 
symmetries. In most of the cases where
a complete set of Yukawa couplings has been allowed
by global symmetries, the models have typically suffered
from a rank one problem in the Yukawa mass matrices, 
rendering them ineffective in giving masses and mixings
to all three generations. This problem may ultimately
be traced to the fact that not all of the intersections
are present on the same torus (in the case of toroidal 
orientifold compactifications). In addition, unlike
heterotic models, the gauge 
couplings are not automatically
unified in intersecting D-brane models.  
Thus, it is non-trivial to find an intersecting 
D-brane model which posesses 
these features.  
Nevertheless, one example is known of an intersecting 
D6-brane model in Type IIA theory on the
$\mathbf{T^6/(\Z_2\times \Z_2)}$ orientifold where these problems may be
solved~\cite{Cvetic:2004ui,Chen:2006gd}. Indeed, this model
automatically realizes tree-level gauge coupling unification,
and it is possible within this model to obtain 
correct Yukawa mass matrices
for the quarks and leptons for specific values of
the moduli VEVs~\cite{Chen:2007px, Chen:2007zu}
\footnote{See references~\cite{Belitsky:2010zr} and~\cite{Lebed:2011zg} for similar models containing 
four generations of chiral fermions.}.

Despite the 
appealing properties of this model, the issue of
moduli stabilization has not yet been fully 
worked out. Although the stabilization of the closed-string 
moduli has already been addressed to 
some extent~\cite{Chen:2006gd, Chen:2007af},  
there is still the problem of 
stabilizing the open-string moduli associated with
D-brane positions in the internal space and Wilson
lines. These unstabilized moduli result in adjoint matter
associated with each stack of D-branes.  Such light-scalars
charged under the SM gauge group are not observed and would have
a detrimental effect on the successful running and eventual
unification of the gauge couplings.    
Moreover, the Yukawa 
couplings for quarks and leptons directly 
depend upon the D-brane positions in 
the internal space as well as other geometric moduli~\cite{Cremades:2003qj}.  
Thus, if one wants to calculate Yukawa couplings, and 
maintain the succesful running of gauge couplings
from the unification scale to the electroweak scale, these
open string moduli must be completely frozen.  

One way to do this is to construct 
intersecting D-brane models where the D-branes 
wrap rigid cycles, which 
was first explored~\cite{Blumenhagen:2005tn} in the
context of Type IIA compactifications on 
$T^6/(\Z_2 \times \Z_2')$ which is 
the only known toroidal background which 
posseses such rigid cycles, as well as the T-dual construction
involving magnetized fractional D-branes~\cite{Dudas:2005jx}.  
The importance
of rigid cycles has also been made more 
clear in recent years by the study of D-instanton
induced superpotential couplings~\cite{Ibanez:2006da,Blumenhagen:2006xt, Florea:2006si}.  
Indeed, such couplings may provide a mechanism for
generating naturally small neutrino masses, 
solving the $\mu$-problem of the MSSM, and 
quite possibly for addressing the issues of supersymmetry 
breaking and inflation.  Moreover, the absence of matter
in the adjoint representation is consistent with heterotic models with a 
$k=1$ Kac-Moody algebra~\cite{Ellis:1990sr}
as well as recent F-theory constructions~\cite{Beasley:2008kw}, 
some of which may be related to Type II vacua by
various chains of dualities.

In this paper, we construct supersymmetric Pati-Salam and MSSM-like models in
the framework of Type IIA flux compactifications
where the D-branes wrap rigid
cycles. This letter is 
organized as follows:  First, we briefly 
review rigid intersecting D6-brane constructions in 
Type IIA on $T^6/(\Z_2 \times \Z_2')$.  We 
then proceed to construct a supersymmetric three-generation
Pati-Salam model in AdS as an example of Type IIA flux vacua. 
We then break the Pati-Salam gauge symmetry to the SM
by displacing the stacks on one torus by requiring them
to run through different fixed points.  
Because the cycles wrapped 
by the D-branes are rigid in these models, the open-string moduli
are completely frozen, and so these fields
will create no difficulities with asymptotic freedom or
with astrophysical constraints on light scalars. 
We do find that it is problematic to break the 
Pati-Salam gauge symmetry to the Standard Model without
the gauge boson of the SM hypercharge becoming massive. 
The SM hypercharge
may remain massless provided that it is extended to 
include $U(1)$'s from other stacks of D-branes.
However, doing
so disrupts the unification of the gauge couplings at
the string scale for the specific model considered.

\section{Intersecting Branes on $T^6/(\Z_2 \times \Z_2')$}

Here, we briefly summarize model building on $\mathbf{T}^6/(\Z_2 \times \Z_2')$.
For a detailed discussion of model building on this background
we direct the reader to~\cite{Blumenhagen:2005tn}, which we 
summarize in the following. 
In Type IIA theory on the $\mathbf{T^6} /(\Z_2 \times \Z_2')$ 
orientifold background, the $\mathbf{T^6}$ is product of three
two-tori and the two orbifold group generators $\theta$, 
$\omega$
act on the complex coordinates $(z_1,z_2,z_3)$ as
\begin{eqnarray}
\theta:(z_1,z_2,z_3)\rightarrow(-z_1,-z_2,z_3) \nonumber \\
\omega:(z_1,z_2,z_3)\rightarrow(z_1,-z_2,-z_3)
\end{eqnarray}
while the antiholomorphic involution $R$ acts as 
\begin{equation}
R(z_1, z_2, z_3)\rightarrow(\bar{z}_1,\bar{z}_2,\bar{z}_3).
\end{equation}
The signs of the $\theta$ action in the $\omega$ sector 
and vice versa have not been specified, and 
the freedom 
to do so is referred to as the choice of discrete torsion.  
One choice of discrete torsion corresponds to the 
Hodge numbers $(h_{11},h_{21}) = (3,51)$ and the other
corresponding to $(h_{11},h_{21}) = (51,3)$.  These 
two different choices are referred to as with discrete 
torsion $(\Z_2 \times \Z_2')$ and without discrete torsion
$(\Z_2 \times \Z_2)$ respectively. For $T^6/(\Z_2 \times \Z_2')$ 
the twisted homology contains collapsed 3-cycles. 
There are 16 fixed points, from which arise 16 
additional 2-cycles with the topology of $\mathbf{P}^1 \cong S^2$.  
As a result, there are 32 collapsed 3-cycles for each twisted sector. 
A $D6$-brane wrapping collapsed 3-cycles in each of the three 
twisted sectors will be unable to move away from a particular 
position on the covering space $\mathbf{T^6}$, and thus the 3-cycle
will be rigid.

A basis of twisted 3-cycles may be defined as
\begin{eqnarray}
[\alpha^{\theta}_{ij,n}] &=& 2[\epsilon^{\theta}_{ij}]\otimes [a^3] \ \ \ \ \ \ \ \ \  [\alpha^{\theta}_{ij,m}] = 2[\epsilon^{\theta}_{ij}]\otimes [b^3],
\end{eqnarray}
\begin{eqnarray}
[\alpha^{\omega}_{ij,n}] &=& 2[\epsilon^{\omega}_{ij}]\otimes [a^1] \ \ \ \ \ \ \ \ \  [\alpha^{\omega}_{ij,m}] = 2[\epsilon^{\omega}_{ij}]\otimes [b^1],
\end{eqnarray}
\begin{eqnarray}
[\alpha^{\theta\omega}_{ij,n}] &=& 2[\epsilon^{\theta\omega}_{ij}]\otimes [a^2] \ \ \ \ \ \ \ \ \  [\alpha^{\theta\omega}_{ij,m}] = 2[\epsilon^{\theta\omega}_{ij}]\otimes [b^2].
\end{eqnarray}
where $[\epsilon^{\theta}_{ij}]$, $[\epsilon^{\omega}_{ij}]$, and $[\epsilon^{\theta\omega}_{ij}]$ denote the 16 fixed points on $\mathbf{T}^2 \times  \mathbf{T}^2$, where $i,j \in {1,2,3,4}$.

A fractional D-brane wrapping both a bulk cycle as well as 
the collapsed cycles may be written in the form
\begin{eqnarray}
\Pi^F_a &=& \frac{1}{4}\Pi^B + \frac{1}{4}\left(\sum_{i,j\in S^a_{\theta}} \epsilon^{\theta}_{a,ij}\Pi^{\theta}_{ij,a}\right)+ \frac{1}{4}\left(\sum_{j,k\in S^a_{\omega}} \epsilon^{\omega}_{a,jk}\Pi^{\omega}_{jk,a}\right)
+ \frac{1}{4}\left(\sum_{i,k\in S^a_{\theta\omega}} \epsilon^{\theta\omega}_{a,ik}\Pi^{\theta\omega}_{ik,a}\right),
\label{fraccycle}
\end{eqnarray}
where the $D6$-brane is required to run 
through the four fixed points for each of 
the twisted sectors.  The set of four fixed points
may be denoted as $S^g$ for the twisted sector $g$.
The constants $\epsilon^{\theta}_{a,ij}$, $\epsilon^{\omega}_{a,jk}$ 
and $\epsilon^{\theta\omega}_{a,ki}$ denote the sign of 
the charge of the fractional brane with respect to the fields 
which are present at the orbifold fixed points.  These signs,
as well as the set of fixed points, must satisfy consistency conditions. 
However, they may be chosen differently for each stack.

The intersection number between a brane $a$ and brane 
$b$ wrapping fractional cycles is 
given by
\begin{eqnarray}
\Pi^F_a \circ \Pi^F_b = \frac{1}{16}[\Pi^B_a \circ \Pi^B_b + 4(n_a^3m_b^3-m_a^3n_b^3)\sum_{i_aj_a\in S^a_{\theta}}\sum_{i_bj_b\in S^b_{\theta}}\epsilon^{\theta}_{a,i_aj_a}\epsilon^{\theta}_{b,i_bj_b}\delta_{i_ai_b}\delta_{j_aj_b} 
+ \\ \nonumber 4(n_a^1m_b^1-m_a^1n_b^1)\sum_{j_ak_a\in S^a_{\omega}}\sum_{j_bk_b\in S^b_{\omega}}\epsilon^{\omega}_{a,j_ak_a}\epsilon^{\omega}_{b,j_bk_b}\delta_{j_aj_b}\delta_{k_ak_b}
+ \\ \nonumber
4(n_a^2m_b^2-m_a^2n_b^2)\sum_{i_ak_a\in S^a_{\theta\omega}}\sum_{i_bk_b \in S^b_{\theta\omega}}\epsilon^{\theta\omega}_{a,i_ak_a}\epsilon^{\theta\omega}_{b,i_bk_b}\delta_{i_ai_b}\delta_{k_ak_b}],
\end{eqnarray}
while the 3-cycle wrapped by the $O6$-plane is given by 
\begin{equation}
\Pi_{O6}=2\eta_{\Omega R}[a^1][a^2][a^3]-2\eta_{\Omega R\theta}[b^1][b^2][a^3]-2\eta_{\Omega R\omega}[a^1][b^2][b^3]-2\eta_{\Omega R\theta\omega}[b^1][a^2][b^3], 
\end{equation}
where the cross-cap charges $\eta_{\Omega R g}$ 
give the RR charge and tension of a given orientifold 
plane $g$, of which there are two
types, $O6^{(-,-)}$ and $O6^{(+,+)}$.  
In this case, $\eta_{\Omega R g} = +1$ indicates 
an $O6^{(-,-)}$ plane, while 
$\eta_{\Omega R g} = -1$ indicates an $O6^{(+,+)}$ 
while the choice of discrete torsion is indicated by the product
\begin{equation}
\eta = \prod_g \eta_{\Omega R g}.
\end{equation}
The choice of no discrete torsion is 
given by $\eta = 1$, while for $\eta = -1$ is 
the case of discrete torsion, for which
an odd number of $O^{(+,+)}$ must be present.

The action of $\Omega R$ on the bulk cycles changes the 
signs of the wrapping numbers as $n^i_a \rightarrow n^i_a$ and $m^i_a \rightarrow -m^i_a$.  However, in addition, there is an action
on the twisted 3 cycle as 
\begin{eqnarray}
\alpha^g_{ij,n} \rightarrow -\eta_{\Omega R}\eta_{\Omega Rg}\alpha^g_{ij,n}, & \alpha^g_{ij,m} \rightarrow \eta_{\Omega R}\eta_{\Omega Rg}\alpha^g_{ij,m}.
\end{eqnarray}
Using these relations, the 
intersection number of a fractional cycle with it's $\Omega R$ image is given by,
\begin{eqnarray}
\Pi'^F_a \circ \Pi^F_a = \eta_{\Omega R}\left(2\eta_{\Omega R}\prod_In^I_am^I_a - 2\eta_{\Omega R\theta}n^3_am^3_a 
-2\eta_{\Omega R\omega}n^1_am^1_a - 2\eta_{\Omega R\theta\omega}n^2_am^2_a\right)
\end{eqnarray}
while the intersection number with the orientifold planes is given by
\begin{eqnarray}
\Pi_{O6} \circ \Pi^F_a = 2\eta_{\Omega R}\prod_I m^I_a - 2\eta_{\Omega R\theta}n^1_an^2_am^3_a - 2\eta_{\Omega R\omega}m^1_an^2_an^3_a - 2\eta_{\Omega R\theta\omega}n^1_am^2_an^3_a.
\end{eqnarray}

\begin{table}[t]
\caption{General spectrum for D6-branes wrapping fractional cycles
and intersecting at generic
angles. ${\cal M}$ represents the multiplicity, and $a_S$ and $a_A$ denote
the symmetric and anti-symmetric representations of
$U(N_a)$, respectively.}
\renewcommand{\arraystretch}{1.4}
\begin{center}
\begin{tabular}{|c|c|}
\hline {\bf Sector} & \phantom{more space inside this box}{\bf
Representation}
\phantom{more space inside this box} \\
\hline\hline
\hline $ab+ba$   & $ {\cal M}(\overline{N}_a,
N_b)=
\Pi^F_a \circ \Pi^F_b$ \\
\hline $a'b+ba'$ & $ {\cal M} (N_a,
N_b)= \Pi^F_{a'} \circ \Pi^F_b $\\
\hline $a'a+aa'$ &  ${\cal M} (a_S)= \Pi^F_{a'} \circ \Pi^F_a - \Pi_{O6} \circ \Pi^F_a$~;~~ ${\cal M} (a_A)=
\Pi^F_{a'} \circ \Pi^F_a + \Pi_{O6} \circ \Pi^F_a $ \\
\hline
\end{tabular}
\end{center}
\label{spectrum}
\end{table}
\noindent The muliticiplity of states in bifundamental, symmetric, and 
antisymmetric representations is shown in Table~\ref{spectrum}.

The fractional cycle wrapped by a D-brane is specified 
by several sets of topological data.  Specifically, the fractional
cycles are described by the bulk wrapping numbers 
$\left\{(n^1, m^1)(n^2, m^2),(n^3, m^3)\right\}$, the sets of fixed points in each of the twisted 
sectors ($S^{\theta}, S^{\omega},~\mbox{and}~S^{\theta\omega}$),
as well as the signs in each twisted sector ($\epsilon^{\theta}_{ij}, \epsilon^{\omega}_{jk},~\mbox{and}~\epsilon^{\theta\omega}_{ki}$). 
Essentially,
the sets of fixed points $S^g$ are specified by the position of the
fractional brane on the three two-tori, while the signs $\epsilon^g_{xy}$ are related
to the choice of discrete Wilson lines for each stack of branes.  
The fixed point sets can in fact be determined for each fractional
brane from the bulk wrapping nubmers.  For a 1-cycle on a $T^2/\Z_2$,
a fractional brane will pass through a pair of fixed points, which can
be determined up to a choice from the wrapping numbers of the 1-cycle:
\begin{equation}
S^{T^2_l} = \left\{\begin{array}{cc}
   \left\{1,~4\right\}~\mbox{or}~\left\{2,~3\right\}\ \ \mbox{for} \ \ (n^l,~m^l) = (\mbox{odd},~\mbox{odd}), \vspace*{0.6cm} \\
\left\{1,~3\right\}~\mbox{or}~\left\{2,~4\right\}\ \ \mbox{for} \ \ (n^l,~m^l) = (\mbox{odd},~\mbox{even}), \vspace*{0.6cm} \\
\left\{1,~2\right\}~\mbox{or}~\left\{3,~4\right\}\ \ \mbox{for} \ \ (n^l,~m^l) = (\mbox{even},~\mbox{odd}),
\end{array}\right.\label{idb:eq:dthdu}
\end{equation}
\noindent
where the two possible choices are related by a transverse
translation of the 1-cycle on the torus.  Let us define the variable
$\delta$ such that $\delta=0$ indicates that we make the first choice, 
while $\delta=1$ indicates the second choice.  For example, 
$(n^l,m^l) = (\mbox{odd},~\mbox{odd})~\mbox{with}~\delta=0~\mbox{indicates}~S^{T^2_l} = \left\{1,~4\right\}$,
while
$(n^l,m^l) = (\mbox{odd},~\mbox{odd})~\mbox{with}~\delta=1~\mbox{indicates}~S^{T^2_l} = \left\{2,~3\right\}$.

From this information, one can then determine the fixed-point sets for each twisted
sector.  This is done by taking the product of the fixed-points sets for each $T^2$
which is acted upon by the orbifold action $g$, i.e. 
\begin{eqnarray}
S^{\theta}_a =& S^{T^2_1}_a\times S^{T^2_2}_a=\left\{i_{a_1} j_{a_1},~i_{a_1} j_{a_2},~i_{a_2} j_{a_1},~
i_{a_2} j_{a_2}\right\}, \nonumber \\
S^{\omega}_a =& S^{T^2_2}_a\times S^{T^2_3}_a=\left\{j_{a_1} k_{a_1},~j_{a_1} k_{a_2},~j_{a_2} k_{a_1},~
j_{a_2} k_{a_2}\right\}, \nonumber \\
S^{\theta\omega}_a =& S^{T^2_3}_a\times S^{T^2_1}_a=\left\{k_{a_1} i_{a_1},~k_{a_1} i_{a_2},~k_{a_2} i_{a_1},~
k_{a_2} i_{a_2}\right\}.
\end{eqnarray}
where $i,~j,~k$ label the pairs of fixed points for each of the three $T^2$ respectively.

\subsection{Twisted Charges and Wilson Lines} 
As stated in Section 2, the signs $\epsilon^{\theta}_{ij,a}$, $\epsilon^{\omega}_{jk,a}$, and $\epsilon^{\theta\omega}_{ki,a}$ are not arbitrary as they must satisfy certain consistency conditions.  Specifically, the set of signs in each twisted sector for each stack of branes $a$ must satisfy
\begin{eqnarray}
\sum_{i,j \in S^a_{\theta}} \epsilon^{\theta}_{a,ij} = 0 \ \ \ \mbox{mod} \ \ \ 4, \\ \nonumber 
\sum_{j,k \in S^a_{\omega}} \epsilon^{\omega}_{a,jk} = 0 \ \ \ \mbox{mod} \ \ \ 4, \\ \nonumber
\sum_{k,i \in S^a_{\theta\omega}} \epsilon^{\theta\omega}_{a,ki} = 0 \ \ \ \mbox{mod} \ \ \ 4,
\end{eqnarray}
and the signs in different twisted sectors for each stack $a$ must be related by the two conditions
\begin{eqnarray}
\epsilon^{\theta}_{a,ij} \epsilon^{\omega}_{a,jk} \epsilon^{\theta\omega}_{a,ki} = 1, \\ \nonumber
\epsilon^{\theta}_{a,ij} \epsilon^{\omega}_{a,jk} = \ \ \mbox{constant} \ \ \forall \ \ j.
\end{eqnarray}

A trivial choice of signs which satisfies these constraints  is just to have them all set to $+1$,  
\begin{equation}
\epsilon^{\theta}_{a,ij} = 1 \ \forall \ ij, \ \ \ \ \ \epsilon^{\omega}_{a,jk} = 1 \ \forall \ jk, \ \ \ \ \ \epsilon^{\theta\omega}_{a,ki} = 1\ \forall \ ki. 
\end{equation}
Another possible non-trivial choice of signs consistent with the constraints is given by
\begin{equation}
\epsilon^{\theta}_{a,ij} = -1 \ \forall \ ij, \ \ \ \ \ \epsilon^{\omega}_{a,jk} = -1 \ \forall \ jk, \ \ \ \ \ \epsilon^{\theta\omega}_{a,ki} = 1\ \forall \ ki.
\end{equation}
More general sets of these signs may be found~\cite{Blumenhagen:2005tn} by setting $\epsilon^{\theta}_{i_1 j_1} = \epsilon^{\theta}_{j_1 k_1}
= \epsilon^{\theta\omega}_{k_1 k_1} = 1$, 
\begin{eqnarray}
\epsilon^{\theta}_{ij} =& \left\{1,~\beta,~\lambda,~\lambda\cdot\beta \right\} \ \ \ \ \ \ \ \ \mbox{for}
\ \ \ \ \ \ \ \ S_{\theta} &= \left\{i_1 j_1,~i_1 j_2,~i_2 j_1,~i_2 j_2\right\}, \\ \nonumber
\epsilon^{\omega}_{jk} =& \left\{1,~\psi,~\beta,~\beta\cdot\psi \right\} \ \ \ \ \ \ \ \ \mbox{for}
\ \ \ \ \ \ \ \ S_{\omega} &= \left\{j_1 k_1,~j_1 k_2,~j_2 k_1,~j_2 k_2\right\}, \\ \nonumber
\epsilon^{\theta\omega}_{ki} =& \left\{1,~\lambda,~\psi,~\lambda\cdot\psi \right\} \ \ \ \ \ \ \ \ \mbox{for}
\ \ \ \ \ \ \ \ S_{\theta} &= \left\{k_1 i_1,~k_1 i_2,~k_2 i_1,~k_2 i_2\right\}, 
\end{eqnarray}
where $\beta,~\lambda,~\mbox{and}~\psi = \pm 1$. The signs $\beta,~\lambda,~\mbox{and}~\psi$
have an interpretation as the choice of discrete Wilson lines along the fractional D-brane.

\subsection{Conditions for Preserving $\mathcal{N}=1$ Supersymmetry}

The condition to preserve $\emph{N}=1$ supersymmetry in four
dimensions is that the rotation angle of any D-brane with respect 
to the orientifold plane is an element of $SU(3)$ \cite{Berkooz:1996km,Cvetic:2001tj}. Essentially, this becomes a constraint on the 
angles made by each stack of branes with respect to the
orientifold planes, \textit{viz}
$\theta^1_a + \theta^2_a + \theta^3_a = 0$ mod $2\pi$, or equivalently  
$\sin(\theta^1_a
+
  \theta^2_a + \theta^3_a)= 0$ and $\cos(\theta^1_a +
  \theta^2_a + \theta^3_a)= 1$.
Applying simple trigonometry, these angles may be expressed in terms of
the wrapping numbers as
\begin{eqnarray}
\tan \theta^i_a=\frac{m^i_a R^i_2}{n^i_a R^i_1} = \frac{m^i_a}{n^i_a}\chi^i
\end{eqnarray}
where $R^i_2$ and $R^i_1$ are the radii of the $i^{\mathrm{th}}$
torus, and $\chi^i = R^i_2/R^i_1$.  We may translate these conditions into restrictions on 
the wrapping numbers as
\begin{eqnarray}
x_A\tilde{A_a}+x_B\tilde{B_a}+x_C\tilde{C_a}+x_D\tilde{D_a}= 0
\label{Eq:SUSY1} \\
A_a/x_A + B_a/x_B + C_a/x_C + D_a/x_D < 0
\label{Eq:SUSY2}
\end{eqnarray}
where we have made the definitions
\begin{eqnarray}
\tilde{A_a} &=& - m^1_am^2_am^3_a, \ \ \ \tilde{B}_a = n^1_an^2_am^3_a, \ \ \ \tilde{C}_a = m^1_an^2_an^3_a, \ \ \ \tilde{D}_a = n^1_am^2_an^3_a, \\
A_a &=& -n^1_an^2_an^3_a, \ \ \ B_a = m^1_am^1_an^3_a, \ \ \ C_a = n^1_am^1_am^3_a, \ \ \  D_a = m^1_an^1_am^3_a.
\end{eqnarray}
and the parameters $x_A$, $x_B$, $x_C$, and $x_D$ are related to 
the complex structure parameters by
\begin{eqnarray}
x_a = \gamma,  \ \ \  x_b = \frac{\gamma}{\chi_2\cdot\chi_3}, \ \ \ x_c = \frac{\gamma}{\chi_1\cdot\chi_3}, \ \ \ \frac{\gamma}{\chi_1\cdot\chi_2}.  
\end{eqnarray}
where $\gamma$ is a positive, real constant.  

\subsection{RR and Torsion Charge Cancellation}
For the present, we focus on supersymmetric AdS vacua with metric, NSNS, and RR fluxes turned on~\cite{Villadoro:2005yq, Camara:2005dc}.  In order to have a consistent model, all RR charges sourced by 
D6-branes, O6-planes, and by the fluxes must
cancel.  
The conditions for the cancellation of RR tadpoles are then given by
\begin{eqnarray}
\sum N_a n_a^1 n_a^2 n_a^3 + \frac{1}{2}(m h_0  + q_1a_1 + q_2a_2 + q_3a_3) =&\ 16 \eta_{\Omega R},  \\ \nonumber
\sum N_a m_a^1 m_a^2 n_a^3 + \frac{1}{2}(m h_1  - q_1b_{11} - q_2b_{21} - q_3b_{31}) =& -16 \eta_{\Omega R\theta}, \\ \nonumber
\sum N_a m_a^1 n_a^2 m_a^3 - \frac{1}{2}(m h_2  - q_1b_{12} - q_2b_{22} - q_3b_{32}) =& -16 \eta_{\Omega R\omega}, & \\ \nonumber
\sum N_a n_a^1 m_a^2 m_a^3 - \frac{1}{2}(m h_3  - q_1b_{13} - q_2b_{23} - q_3b_{33}) =& -16 \eta_{\Omega R\theta}.
\end{eqnarray}
where $a_i$ and $b_{ij}$ arise due to the metric fluxes, $h_0$ and $h_i$ arise due
to the NSNS fluxes, and $m$ and $q_i$ arise from the RR fluxes. We consider these fluxes
to be quantized in units of eight in order to avoid subtle problems with Dirac flux
quantization conditions.  

For simplicity, we set all of the K\a"ahler moduli equal to each other, $T_1 = T_2 = T_3=T$,
so that we then obtain $q_1 = q_2 = q_3=q$ from the superpotential.  In order to satisfy
the Jacobi identities for the metric fluxes, we shall consider the solution 
$a_i = a$, $b_{ii} = -b_i$, and $b_{ji} = b_i$, where $j \neq i$.  

In order to obtain supersymmetric minima, it must be required that
\begin{equation}
3 a \mbox{Re}(S) = b_i \mbox{Re}(U^i),
\label{Eq:SUSYmin}
\end{equation}  
where
\begin{equation}
\mbox{Re}(S) = \frac{e^{-\phi_4}}{\sqrt{\chi^1\chi^2\chi^3}}, \ \ \mbox{Re}(U^i) = e^{-\phi_4}\sqrt{\frac{\chi^j\chi^k}{\chi^i}} \ \ \ \ \ i \neq j \neq k,
\end{equation}
with $S$ and $U^i$ being the dilaton and complex structure moduli respectively, and
where $\phi_4$ is the four-dimensional dilaton.  Then, we have the relations
\begin{equation}
b_1 = \frac{3 a}{\chi_2 \chi_3}, \ \ \ \ \ b_2 = \frac{3 a}{\chi_1 \chi_3}, \ \ \ \ \ 
b_3 = \frac{3 a}{\chi_1 \chi_2}.
\end{equation}
In addition, there are consistency conditions which must be satisfied
\begin{equation}
3 h_i a + h_0 b_i = 0, \ \ \ \ \ \mbox{for} \ \ \ i = 1,~2,~3,
\label{Eq:ConCon}
\end{equation}
so that we then have
\begin{equation}
h_1 = -\frac{h_0}{\chi_2 \chi_3}, \ \ \ \ \ h_2 = -\frac{h_0}{\chi_1\chi_3}, \ \ \ \ \ 
h_3 = -\frac{h_0}{\chi_1 \chi_2}.
\end{equation}
Thus, the RR tadpole conditions can be written in a simplified form as
\begin{eqnarray}
\label{bulktadpole}
\sum N_a n_a^1 n_a^2 n_a^3 + \frac{1}{2}(m h_0  + 3 a q) =&\ 16 \eta_{\Omega R},  \\ \nonumber
\sum N_a m_a^1 m_a^2 n_a^3 - \frac{1}{2\chi_2\chi_3}(m h_0  + 3 a q) =& -16 \eta_{\Omega R\theta}, \\ \nonumber
\sum N_a m_a^1 n_a^2 m_a^3 - \frac{1}{2\chi_1\chi3}(m h_0  + 3 a q) =& -16 \eta_{\Omega R\omega}, & \\ \nonumber
\sum N_a n_a^1 m_a^2 m_a^3 - \frac{1}{2\chi_1\chi_2}(m h_0  + 3 a q) =& -16 \eta_{\Omega R\theta}.
\end{eqnarray}

Since $(m h_0  + 3 a q)$ can be either positive or negative, the supergravity fluxes
can contribute either positive or negative D6-brane charge.  Therefore, since we may
also have an odd number of $O6^{++}$ planes as well as hidden sector branes, the 
RR-tadpole conditions are somewhat relaxed.  However, we are still constrained
by the requirement of torsion charge cancellation.  Finally, it can be shown~\cite{Camara:2005dc} that if Eqs.~(\ref{Eq:SUSY1}),(~\ref{Eq:SUSYmin}), and (\ref{Eq:ConCon}) are satisfied,
then the conditions for the Freed-Witten anomalies to be cancelled,
\begin{equation}
h_0\tilde{A}_a + h_1\tilde{B}_a + h_2\tilde{C}_a + h_3\tilde{D}_a = 0
\end{equation}
are automatically satisified.

In order to ensure torsion charge cancellation, me must satisfy
\begin{eqnarray}
\sum_a N_a n_a^i(\epsilon^i_{a,kl} - \eta_{\Omega R}\eta_{\Omega R i} \epsilon^i_{a,kl}),
\\ 
\sum_a N_a m_a^i(\epsilon^i_{a,kl} + \eta_{\Omega R}\eta_{\Omega R i} \epsilon^i_{a,kl}),
\end{eqnarray}
where the sums are over \textit{each} each fixed point $[e^g_{ij}]$. Clearly, these 
conditions are non-trivial to be satisfied.

\subsection{The Green-Schwarz Mechanism}
Although the total non-Abelian anomalies cancel automatically when
the RR-tadpole conditions are satisfied, additional mixed
anomalies like the mixed gravitational anomalies which generate
massive fields are not trivially zero \cite{Cvetic:2001tj}. 
 These anomalies are cancelled by a generalized
Green-Schwarz (G-S) mechanism which involves untwisted
Ramond-Ramond forms. Integrating the G-S couplings of the
untwisted RR forms to the $U(1)$ field strength $F_a$ over the
untwisted cycles of $\mathbf{T^6/(\Z_2\times \Z'_2)}$ orientifold, we find
\begin{eqnarray}
\int_{D6^{untw}_a} C_5 \wedge \textrm{tr}F_a \sim N_a \sum_i
r_{ai}\int_{M_4} B^i_2 \wedge \textrm{tr}F_a,
\end{eqnarray}
where
\begin{equation}
B^i_2 = \int_{[\Sigma_i]} C_5,\;\; [\Pi_a]=\sum^{b_3}_{i=1}
r_{ai}[\Sigma_i],
\end{equation}
and
${[\Sigma_i]}$ is the basis of homology 3-cycles, $b_3=8$. Under orientifold action 
only half survive.  In other words,
$\{r_{ai}\}=\{\tilde{B}_a, \tilde{C}_a, \tilde{D}_a,
\tilde{A}_a\}$ in this definition. Thus the couplings of the four
untwisted RR forms $B^i_2$ to the $U(1)$ field strength $F_a$ are
\cite{Aldazabal:2000dg}
\begin{eqnarray}
  N_a \tilde{B}_a \int_{M_4}B^1_2\wedge \textrm{tr}F_a,&&  \;
  N_a \tilde{C}_a \int_{M_4}B^2_2\wedge \textrm{tr}F_a,
   \nonumber \\
  N_a \tilde{D}_a \int_{M_4}B^3_2\wedge \textrm{tr}F_a,&&  \;
  N_a \tilde{A}_a \int_{M_4}B^4_2\wedge \textrm{tr}F_a.
\end{eqnarray}

Besides the contribution to G-S mechanism from untwisted 3-cycles,
the contribution from the twisted cycles should be taken into
account. As in the untwisted case we integrate the Chern-Simons coupling over the
exceptional 3-cycles from the twisted sector.  We choose the sizes
of the 2-cycles on the topology of $S^2$ on the orbifold
singularities to make the integrals on equal foot to those from
the untwisted sector. Consider the twisted sector $\theta$ as
an example,
\begin{eqnarray}
\int_{D6^{tw,\theta}_a}C_5\wedge {\rm tr}F_a \sim   N_a
\sum_{i,j\in S^a_{\theta}} \epsilon^{\theta}_{a,ij} m^3_a
\int_{M_4} B^{\theta ij}_2 \wedge {\rm tr}F_a,
\end{eqnarray}
where $B^{\theta ij}_2=\int_{[\alpha^{\theta}_{ij,m}]}C_5$, with
orientifold action taken again and with the choice of crosscap
charges
$\eta_R = -\eta_{R\theta} = -\eta_{R\omega} = -\eta_{R\theta\omega} = -1$.
Although $i,j$ can run through $\left\{1-4\right\}$ for each
of the four fixed points in each sector, these are constrained
by the wrapping numbers from the untwisted sector so that only four
possibilities remain.  A similar argument may be applied for $\omega$ and 
$\theta\omega$ twisted sectors:
\begin{eqnarray}
\int_{D6^{tw,\omega}_a}C_5\wedge {\rm tr}F_a \sim   N_a
\sum_{j,k\in S^a_{\omega}} \epsilon^{\omega}_{a,jk} m^1_a
\int_{M_4} B^{\omega jk}_2 \wedge {\rm tr}F_a.
\end{eqnarray}
\begin{eqnarray}
\int_{D6^{tw,\theta\omega}_a}C_5\wedge {\rm tr}F_a \sim   N_a
\sum_{i,j\in S^a_{\theta\omega}} \epsilon^{\theta\omega}_{a,ik}
m^2_a \int_{M_4} B^{\theta\omega ik}_2 \wedge {\rm tr}F_a.
\end{eqnarray}

In summary, there are twelve additional couplings of the
Ramond-Ramond 2-forms $B^i_2$ to the $U(1)$ field strength $F_a$
from the twisted cycles, giving rise to massive $U(1)$'s.  However
from the consistency condition of the $\epsilon$'s (see section
3.1) related  to the
discrete Wilson lines they may be dependent or degenerate.  So
even including the couplings from the untwisted sector we still
have an opportunity to find a linear combination for a massless
$U(1)$ group.  Let us write down these couplings of the twisted
sector explcitly:
\begin{eqnarray}
N_a  \epsilon^{\theta}_{a,ij} m^3_a
\int_{M_4} B^{\theta ij}_2 \wedge {\rm tr}F_a, \ \ \ 
N_a  \epsilon^{\omega}_{a,jk} m^1_a
\int_{M_4} B^{\omega jk}_2 \wedge {\rm tr}F_a, \nonumber \\
N_a  \epsilon^{\theta\omega}_{a,ik} m^2_a \int_{M_4}
B^{\theta\omega ik}_2 \wedge {\rm tr}F_a.
\end{eqnarray}

Checking the mixed cubic anomaly by introducing the dual field of
$B^i_2$ in the diagram, we can find the contribution from both
untwisted and twisted sectors having a intersection number form
and which is cancelled by the RR-tadpole conditions mentioned.
These couplings determine the linear combinations of $U(1)$ gauge
bosons that acquire string scale masses via the G-S mechanism.  Thus, in constructing
MSSM-like models, we
must ensure that the gauge boson of the hypercharge $U(1)_Y$ group
does not receive such a mass. In general, the hypercharge is a linear combination
of the various $U(1)$s generated from each stack :
\begin{equation}
U(1)_Y=\sum_a c_a U(1)_a.
\end{equation}
The corresponding field strength must be orthogonal to those that
acquire G-S mass.  Thus we demand 
\begin{eqnarray}
\sum_a c_a N_a 
\epsilon^{\omega}_{a,jk} m^1_a&=& 0, \ \ \ \ 
\sum_a c_a N_a
\epsilon^{\theta\omega}_{a,ki} m^2_a = 0, \ \ \ \ 
\sum_a c_a N_a
\epsilon^{\theta}_{a,ij} m^3_a = 0, 
\label{twistedGS}
\end{eqnarray}
for the twisted couplings as well as 
\begin{eqnarray}
 \sum_a c_a N_a \tilde{A_a} &=& 0, \ \ \ \ 
 \sum_a c_a N_a \tilde{B_a}  = 0, \ \ \ \ 
 \sum_a c_a N_a \tilde{C_a}  = 0, \ \ \ \ 
 \sum_a c_a N_a \tilde{D_a}  = 0, 
\label{GSeq}
\end{eqnarray}
for the untwisted.

\section{Model Building}

It is well-known that intersecting D-brane models with a
Pati-Salam gauge group are the only models where 
it is possible to have all Yukawa couplings for quarks and 
leptons present at the stringy tree-level.  In the case of 
$T^6/(\Z_2 \times \Z_2')$ where all D-branes are wrapping rigid cycles, 
Pati-Salam models are also favored due to both the twisted charge cancellation
conditions as well as the K-theory consistency conditions.  

D-brane instantons have also been greatly studied in recent years and may 
play an important role in generating nonperturbative contributions to the superpotential.
In particular, they may in principle generate couplings which are perturbatively
forbidden by global symmetries which have their origin in $U(1)$ gauge factors
which become massive via a generalized Green-Schwarz mechanism.  In order
to have the proper zero-mode structure, the Euclidean 2-branes giving rise
to such instantons must wrap rigid-cycles.  The only known toroidal orientifold
where rigid cycles are available is none other than $T^6/(\Z_2 \times \Z_2')$.  

Although the potential to generate perturbatively forbidden superpotential couplings
via D-brane instanton effects is exciting, it is in fact non-trivial to satisfy all
consistency conditions required for them to be incorporated.  Besides the aforementioned
restriction that the E2-branes wrap rigid cycles, it is also necessary for the cycles wrapped
by the E2-brane to be invariant under the orientifold projection which strongly restricts
the possible cycles in a way which depends on the choice of cross-cap charges.  In addition,
the E2-brane must have the correct intersection numbers with the relevant D-branes in order 
to generate a potential coupling, and may not intersect any other D-branes.  Clearly, this is
a very difficult condition to satisfy.  For the models we will construct, we will attempt 
to cancel the twisted charges by using stacks of branes wrapping cycles whose 
bulk component is invariant under the orientifold action.  Our motivation in doing this
is to try to make it less likely that any E2 brane that we may consider in the model
will have additional intersections with these stacks, a necesary requirement to
generate required couplings such as a $\mu$-term or a neutrino Majorana mass term.

\subsection{A Three-family Pati-Salam Model}

\begin{table}[tf]
\footnotesize
\renewcommand{\arraystretch}{1.0}
\caption{A set of D6-brane configurations for a three-family Pati-Salam model
in Type IIA on the $\mathbf{T}^6 /(\Z_2 \times \Z_2')$
orientifold, where the D6-branes are wrapping rigid cycles. This configuration preserves $\mathcal{N}=1$ supersymmetry for $\chi_1=1$, $\chi_2=2$, and $\chi_3=1$. The bulk tadpole conditions Eqn.~\ref{bulktadpole} are satisfied for this model by choosing $(m h_0  + 3 a q) = -64$, and all twisted tadpoles are cancelled.}
\label{rigidmodel3gen}
\begin{center}
\begin{tabular}{|c||c|c||c|c|c|}
\hline
  $N$ & \mbox{stack} & $(n^1,m^1)\times (n^2,m^2)\times

(n^3,m^3)$ & $(\beta,~\lambda,~\psi)$ & 
 $(\delta_1,~\delta_2,~\delta_3)$\\

\hline

     4 & a & $(-1,-1)\times (~0,~1)\times (~1,~2)$ & (~1,~1,~1) & (~1,~1,~1)\\
    
     2 & b & $(~1,-2)\times (~0,-1)\times (-1,~1)$ & (~1,~1,~1) & (~1,~1,~1)\\
     
     2 & c & $(-1,~2)\times (~0,-1)\times (~1,-1)$ & (~1,~1,~1) & (~1,~1,~1)\\

\hline
\hline

     4 & $\alpha_1$ & $(~1,~1)\times (~0,~1)\times (-1,-2)$ & (~1,-1,~1) & (~1,~1,~1)\\

     4 & $\alpha_2$ & $(~0,~1)\times (~0,~1)\times (-1,~0)$ & (~1,-1,-1)& (~1,~1,~1)\\
     
     4 & $\alpha_3$ & $(~0,~1)\times (~0,~1)\times (-1,~0)$ & (~1,~1,-1)& (~1,~1,~1)\\    
    
\hline
\hline
  
      4 & $\beta_1$ & $(~1,-1)\times (~1,~0)\times (~1,~1)$ & \mbox{bulk} & \mbox{bulk} \\
      
      6 & $\beta_2$ & $(~1,~0)\times (~2,-1)\times (~1,~1)$ & \mbox{bulk} & \mbox{bulk} \\
     
     22 & 1 & $(~1,~0)\times (~0,-1)\times (~1,~1)$ & \mbox{bulk} & \mbox{bulk} \\
     
     76 & 2 & $(~0,-1)\times (~1,~0)\times (~0,~1)$ & \mbox{bulk} & \mbox{bulk} \\
     
     24 & 3 & $(~0,-1)\times (~0,~1)\times (~1,~0)$ & \mbox{bulk}&  \mbox{bulk} \\
\hline          

\end{tabular}
\end{center}
\end{table}

A simple way to cancel twisted tadpoles is to construct models
models where the stacks of branes giving rise to the observable sector all wrap bulk
cycles which are homologically the same, but which differ in their twisted cycles, such
as the model considered in~\cite{Chen:2006sd}.  Models with a Pati-Salam gauge group are particularly
well-suited for this type of construction.  Besides
making it easier to cancel twisted tadpoles and satisfy K-theory constraints, the gauge couplings are automatically unified with a canonical
normalization in such types of models. However, such constructions tend to result in four-family models. 

A slight variation on this idea is to have stacks of 
branes which are not technically the same homologically, but which are related by some interchange
symmetry.   
The wrapping numbers and twisted charge assignments
for a model of this type are shown in Table~\ref{rigidmodel3gen}, where we have made the choice of 
discrete torsion $\eta_{\Omega R} = -\eta_{\Omega R\theta} = -\eta_{\Omega R\omega} = -\eta_{\Omega R\theta\omega} = -1$. 
\begin{table}[tb]
\footnotesize
\renewcommand{\arraystretch}{1.0}
\caption{Intersection numbers for the Pati-Salam model with the D6-brane configurations shown in Table~\ref{rigidmodel3gen}.}
\label{MI-Numbers}
\begin{center}
\begin{tabular}{|c||c|c||c|c|c|c|c|c|c|c|c|c|c|c|c|c|c|c|c|c|}
\hline
& \multicolumn{20}{c|}{$SU(4)_C\times SU(2)_L\times SU(2)_R \times SU(4)^4 \times SU(6) \times \prod_{i=1}^3 {USp(N_i)}$}\\
\hline \hline  & $N$ & $n_{S}$& $n_{A}$ & $b$ & $b'$ & $c$ & $c'$&  $\alpha_1$ & $\alpha_1'$ & $\alpha_2$ & $\alpha_2'$ & $\alpha_3$ & $\alpha_3'$ & $\beta_1$ & $\beta_1'$ & $\beta_2$ & $\beta_2'$ & 1 & 2 & 3  \\

\hline

    $a$&  4&  0 &  6  & 3 & 0 & -3 & 0 & 2 & 0 & 1 & 0 & -2 & 0 & 2 & 0 & 2 & -6 & 0 & 0 & 0 \\

    $b$&  2&  0 & -6  & - & - & 0 & -6 & 3 & 0 & 0 & 0 & 1 & 0 & -2 & 0 & -8 & 0 & 0 & 0 & 0 \\

    $c$&  2&  0 & -6  & - & - & - & - & -3 & 0 & 0 & 0 & -1 & 0 & -2 & 0 & -8 & 0 & 0 & 0 & 0 \\

\hline

    $\alpha_1$&  4&  0 & 6  & - & - & - & - & - & - & 1 & 1 & 2 & 0 & 2 & 0 & 2 & -6 & 0 & 0 & 0 \\
    
    $\alpha_2$&  4&  0 & 0  & - & - & - & - & - & - & - & - & 0 & 0 & -1 & -1 & -2 & -2 & 0 & 0 & 0\\
    
    $\alpha_3$&  4&  0 & 0  & - & - & - & - & - & - & - & - & - & - & -1 & -1 & -2 & -2 & 0 & 0 & 0\\

\hline
    $\beta_1$ &   4 & 0 & 0   & \multicolumn{17}{c|}{$\chi_1=1/\sqrt{2},~
\chi_2=\sqrt{2},~\chi_3=1/\sqrt{2}$}\\

    $\beta_2$ &   6 & 0  & 0  & \multicolumn{17}{c|}{}\\

    1&   22& 0  & 0  & \multicolumn{17}{c|}{}\\

    2&   76& 0 & 0    & \multicolumn{17}{c|}{}\\

    3&   24&  0 & 0   & \multicolumn{17}{c|}{$m h_0  + 3 a q = -64$}\\

\hline

\end{tabular}

\end{center}

\end{table}

Stacks $a$, $b$, and $c$ comprise the \lq observable\rq \ sector,
which results in a three-family model with Pati-Salam gauge group.  
Interestingly, the bulk intersection numbers between stack $a$ and stacks
$b$ and $c$ respectively are zero, which would result in a strictly  non-chiral
spectrum if these stacks were wrapping bulk cycles.  However, one-half of 
each vector pair is projected out by the twisted actions, resulting in a net
chiral spectrum. 

The additional
stacks $\alpha_i,~i=1-3$ wrapping rigid cycles are present in order to satisfy the twisted tadpole
conditions, while the stacks $1-3$ wrapping bulk cycles are present
in order to satisfy the untwisted tadpole conditions.  In order to preserve $\mathcal{N}=1$
supersymmetry with these sets of stacks, we only need to require that the structure parameters satisfy
$x_a = 2x_c$.  The parameters $x_b$ and $x_d$ are completely arbitrary due symmetries
relating the stacks. In order to also fix $x_b$ and $x_c$ in terms of $x_a$, let us 
also introduce the stacks labeled $\beta_1$ and $\beta_2$ into the model. We should emphasize that there is some freedom in choosing these stacks so that the particular configuration that we have chosen is just one possible solution.  With these additions, we must set $x_a = x_b = 2x_c = x_d$ in order to preserve
supersymmetry, from which we have
\begin{equation}
\chi_1 = 1/\sqrt{2}, \ \ \ \ \ \chi_2 = \sqrt{2}, \ \ \ \ \ \chi_3 = 1/\sqrt{2}.
\end{equation}
With this configuration, the tadpole conditions may be satisfied by setting
\begin{equation}
m h_0  + 3 a q = -64.
\end{equation}

The intersection numbers between the stacks and their images respectively are shown in Table~\ref{MI-Numbers}.
The resulting matter spectrum is shown in Table~\ref{PSspectrum}.  
One unsatisfactory aspect of this model is that there exist states which are charged
under both the observable and hidden sector gauge groups, a generic problem in these types of models. In an ideal model, these two sectors
should either be completely sequestered from one another or all states charged under both sectors should
become heavy enough to evade experimental constraints.  However, it is clear that 
there are actually many possible hidden sectors which may be compatible with the observable sector of this model.  
In fact, it is likely that there is a \lq landscape\rq \ of hidden sectors, whereas that shown
in Table~\ref{rigidmodel3gen} is but one solution.  It is quite possible that there is at 
least one hidden sector somewhere in this landscape that could satisfy all phenomenological constraints.  
Indeed, it is possible to take the rank of the hidden sector groups to be very large and at the same time turning on a
large amount of flux in such a way as to ensure both twisted and untwisted tadpole cancellation.  
We may eliminate the adjoint fields associated with these stacks by requiring them
to wrap rigid cycles so that these gauge groups will have large and negative beta functions.  The exotic
matter charged under both observable and hidden sector gauge groups can then become quite massive since
these groups can be made to become confining at high energy scales.  

In order for the gauge boson of a $U(1)$ factor to remain massless, it must satisfy the Green-Schwarz 
anomaly cancellation conditions Eqs.~(\ref{GSeq}) and (\ref{twistedGS}).  For this Pati-Salam model,
it turns out that none of the $U(1)$ groups remain massless, and the effective gauge symmetry is
\begin{equation}
SU(4)_C\times SU(2)_R\times
SU(2)_L \times SU(4)^4\times SU(6) \times \prod_{i=1}^3 Usp(N_i).
\end{equation}
These $U(1)$ groups remain as global symmetries which may perturbatively forbid desirable couplings.  However, 
these global symmetries may be broken by D-brane instanton effects.

\begin{table}[tb]

\footnotesize

\renewcommand{\arraystretch}{1.0}

\caption{The chiral and vector-like superfields, and their quantum
numbers under the gauge symmetry $SU(4)_C\times SU(2)_R\times
SU(2)_L \times SU(4)^4 \times SU(6) \times \prod_{i=1}^3 Usp(N_i)$.}

\label{PSspectrum}

\begin{center}

\begin{tabular}{|c||c||c|c|c||c|c|c|}\hline

 & Quantum Number

& $Q_4$ & $Q_{2R}$ & $Q_{2L}$  & Field \\

\hline\hline

$ab$ & $3\times (\overline{4},2,1,1,1,1,1,1,1,1,1)$ & -1 & 1 & $0$   & $F_R(Q_R, L_R)$\\

$ac$ & $3 \times (4,1,\overline{2},1,1,1,1,1,1,1,1)$ & 1 & 0 & -1  & $F_L(Q_L, L_L)$\\

$a_{A}$ & $6\times(6,1,1,1,1,1,1,1,1,1,1)$ & 2 & 0 & 0   & $D_1(D_1^c, D_1)$ \\

$b_{A}$ & $6\times(1,\overline{1},1,1,1,1,1,1,1,1,1)$ & 0 & -2 & 0   & $S_R^i$ \\

$c_{A}$ & $6\times(1,1,\overline{1},1,1,1,1,1,1,1,1)$ & 0 & 0 & -2   & $S_L^i$ \\

\hline

$bc$ & $6 \times (1,\overline{2},2,1,1,1,1,1,1,1,1)$ & 0 & -1 & 1   &  $H_u^i$, $H_d^i$\\

& $6 \times (1,2,\overline{2},1,1,1,1,1,1,1,1)$ & 0 & 1 & -1   & $\overline{H}_u^i$, $\overline{H}_d^i$ \\ 

\hline

$b'c$ & $6 \times (1,2,2),1,1,1,1,1,1,1,1)$ & 0 & 1 & 1   & $H_1(H_1^i$, $H_2^i)$\\

\hline
${\alpha_{1}}_{A}$ & $6\times(1,1,1,6,1,1,1,1,1,1,1)$ & 0 & 0 & 0   & - \\

$a'\alpha_1$ & $2 \times (\overline{4},1,1,\overline{4},1,1,1,1,1,1,1)$ & -1 & 0 & 0  & - \\

$a'\alpha_2$ & $1 \times (\overline{4},1,1,1,\overline{4},1,1,1,1,1,1)$ & -1 & 0 & 0  & - \\

$a'\alpha_3$ & $2 \times (\overline{4},1,1,1,1,\overline{4},1,1,1,1,1)$ & -1 & 0 & 0  & - \\

$b\alpha_1$ & $3 \times (1,\overline{2},1,4,1,1,1,1,1,1,1)$ & 0 & -1 & 0  & - \\

$b\alpha_3$ & $1 \times (1,\overline{2},1,1,1,4,1,1,1,1,1)$ & 0 & -1 & 0  & - \\

$c\alpha_1$ & $1 \times (1,1,2,\overline{4},1,1,1,1,1,1,1)$ & 0& 0 & 1 & - \\

$c\alpha_3$ & $1 \times (1,1,2,1,1,\overline{4},1,1,1,1,1)$ & 0& 0 & 1 & - \\

$\alpha_1\alpha_2$ & $1 \times (1,1,1,\overline{4},4,,1,1,1,1,1)$ & 0 & 0 & 0  & - \\

$\alpha_1'\alpha_2$ & $1 \times (1,1,1,\overline{4},\overline{4},,1,1,1,1,1)$ & 0 & 0 & 0  & - \\

\hline

\end{tabular}

\end{center}

\end{table}

Although the adjoint fields have been eliminated by splitting the bulk D-branes into their fractional consituents, light non-chiral matter in the
bifundamental representation may still appear between pairs of fractional branes~\cite{Blumenhagen:2005tn}. 
These non-chiral states smoothly connect the configuration of fractional D-branes to one consisting of non-rigid D-branes. 
In the present case, stacks $b$ and $c$ are wrapping bulk cycles which are homologically identical, but differ in their twisted cycles. 
Thus, the required states to play the role of the Higgs fields,
$H_u^i$, $H_d^i$ are present in $bc$ sector in the form of non-chiral matter.  The Higgs states arising from this sector were also
present in the model discussed in~\cite{Chen:2006sd}. Although the Higgs states required to give masses and quarks and leptons may arise in the non-chiral sector, there are additional Higgs-like states, $H_1^i$, $H_2^i$, in the $b'c$ sector. Although these states are charged under global $U(1)$ factors which would forbid the Yukawa couplings involving these states, it is possible that these couplings may be induced via D-brane instanton effects which can violate such global $U(1)$ symmetries.

\subsection{D-brane Instanton Induced Superpotential Couplings}

Non-perturbative D-brane instanton induced
superpotential couplings have been receiving a great
 deal of attention of late. In particular, superpotential
couplings such as a Majorana mass term for right-handed neutrinos
as well as the $\mu$-term for the Higgs states are typically
forbidden perturbatively due to selection rules which arise
from global symmetries.  These global symmetries arise
due to the $U(1)$ gauge groups associated which each stack of
D-branes whose gauge bosons pick up string-scale masses via a
generalized Green-Schwarz mechanism.  Under suitable conditions,
Euclidean D2-brane (E2) instantons may break these
global $U(1)$ symmetries and generate $U(1)$ violating
interactions.

An E2-brane wrapping a cycle $\Xi$ which intersects a stack
of D6-branes $a$ wrapping a cycle $\Pi_a$ will result in
zero-modes present at the intersection,
which results in the E2-brane being charged under the $U(1)_a$ gauge group,
\begin{equation}
Q_a = N_a \Xi \circ (\Pi_a - \Pi_a').
\label{instanton1}
\end{equation}
Perturbatively forbidden superpotential couplings may
by then be generated by an E2-instanton provided it
has a suitable zero-mode structure.  In particular,
the cycle $\Xi$ must be rigid to ensure that there
exist no reparametrization zero modes.  In addition,
the cycle wrapped by the instanton must not have
an intersection with its own image, $\Xi \cap \Xi' = 0$,
or equivalently the cycle wrapped by the instanton must
be invariant under the orientifold projection and carry
gauge group $O(1)$, in which case Eq.~(\ref{instanton1}) is modified
to
\begin{equation}
Q_a = -N_a \Xi \circ \Pi_a.
\end{equation}

\begin{table}[t]
\footnotesize
\renewcommand{\arraystretch}{0.75}
\caption{Cycles wrapped by an E2 instanton.}
\label{instanton}
\begin{center}
\begin{tabular}{|c||c|c||c|c|c|}
\hline
 E2 & $N$ & $(n^1,l^1)\times (n^2,l^2)\times

(n^3,l^3)$ & $(\beta,~\lambda,~\psi)$ &
 $(\delta_1,~\delta_2,~\delta_3)$\\

\hline

   $E2_1$  & 1 & $(-1,~0)\times (-1,~0)\times (1,~0)$ & (~1,~1,~1) & (~0,~0,~0)\\
   
\hline

\end{tabular}

\end{center}

\end{table}

Let us consider the E2-branes wrapping the fractional cycles shown in Table~\ref{instanton}, 
with the intersection numbers shown in Table~\ref{InstaintNumbers}.  With the zero mode structure of $E2_1$, we 
may induce a Yukawa term in the superpotential of the form
\begin{equation}
W_Y = c F_L F_R H_1, 
\end{equation}
which can provide masses for the quarks and leptons.   Thus, it may be possible to obtain 
the mass hierarchies for quarks and leptons from these instanton induced couplings.  We defer a deeper analysis
of this question to future work.  

\begin{table}[t]
\footnotesize
\renewcommand{\arraystretch}{1.0}
\caption{Intersection numbers of E2 branes with the D6-brane configurations shown in Table~\ref{rigidmodel3gen}.}
\label{InstaintNumbers}
\begin{center}
\begin{tabular}{|c||c|c|c|c|c|c|c|c|c|c|c|c|c|}
\hline

\hline  E2 & $a$ &  $b$ &  $c$ &  $\alpha_1$  & $\alpha_2$  & $\alpha_3$  & $\beta_1$ & $\beta_1'$ & $\beta_2$ & $\beta_2'$ & 1 & 2 & 3 \\

\hline

    $E2_1$&      0   &   1  &   0  &    0         &   0         &  0         &  0 & 0 & 0 & 0 & 0 & 0 & 0 \\   
\hline

\end{tabular}

\end{center}

\end{table}

\section{A Three-generation MSSM-like Model}
The Pati-Salam model of the previous section may be broken to the Standard Model explicitely by splitting the stacks on one torus such that 
\begin{equation}
a \rightarrow a1~+~a2 \ \ \ \ \  \mbox{and} \ \ \ \ \ b \rightarrow b1~+~b2
\end{equation}
where $N_{a1} = 3$, $N_{a2} = 1$,
$N_{b1} = 1$ and $N_{b1} = 1$ as shown in Table~\ref{rigidmodel3gen_MSSM}. For D6-branes wrapping non-rigid
cycles, this would be accomplished by giving a VEV to an adjoint scalar associated with each stack.  However,
for the present model, the adjoint scalars are not present since all D6-branes are wrapping rigid cycles.  Thus,
to split the stacks, we simply require that stacks $a1$ and $a2$, as well as $c1$ and $c2$, must pass through different
fixed points on the third torus recalling from before that for a given set of wrapping numbers, there are two
possible choices for the fixed points in which a particular one-cycle will pass.  These stacks will then wrap
different fractional cycles, thus breaking the gauge symmetry.   Since the D6-brane configuration has been changed slightly from the previous
Pati-Salam model, the twisted charge cancellation will also be modified.  Indeed, we must add an additional stack(s) in order to cancel the twisted tadpoles, as can be seen in Table~\ref{rigidmodel3gen_MSSM} where stacks $d_i$, $i=1-4$ and $e_j$, $j=1-8$ have been added to the previous Pati-Salam model.  With this configuration of D6-branes, all consistency conditions are satisfied and  $\mathcal{N}=1$ supersymmetry is preserved.  
In general, one would not expect to have the same intersection numbers for two stacks which have been split in this way.  Remarkably, for the present model, the number of Standard Model fermions remains three even after splitting
the stacks. The resulting spectrum is then that of a three-family MSSM model with an extended gauge group
where the hypercharge $U(1)_Y$ is defined by Eq.~(55).  The intersection numbers are shown
in Table~\ref{SMintnum} and the corresponding MSSM matter spectrum is shown in Table~\ref{MSSMspec}. For brevity, we do not show the exotic states charged under
both observable and hidden sector gauge groups.  

\begin{table}[pf]
\footnotesize
\renewcommand{\arraystretch}{1.0}
\caption{A set of D6-brane configurations for a three-generation MSSM-like model
in Type IIA on the $\mathbf{T}^6 /(\Z_2 \times \Z_2')$
orientifold, where the D6-branes are wrapping rigid cycles. This configuration preserves $\mathcal{N}=1$ supersymmetry for $\chi_1=1$, $\chi_2=2$, and $\chi_3=1$. The bulk tadpole conditions Eq.~(\ref{bulktadpole}) are satisfied for this model by choosing $(m h_0  + 3 a q) = -256$, and all twisted tadpoles are cancelled.}
\label{rigidmodel3gen_MSSM}
\begin{center}
\begin{tabular}{|c||c|c||c|c|c|}
\hline
  $N$ & \mbox{stack} & $(n^1,m^1)\times (n^2,m^2)\times

(n^3,m^3)$ & $(\beta,~\lambda,~\psi)$ & 
 $(\delta_1,~\delta_2,~\delta_3)$\\

\hline

     3 & a1 & $(-1,-1)\times (~0,~1)\times (~1,~2)$ & (~1,~1,~1) & (~1,~1,~1)\\
     
     1 & a2 & $(-1,-1)\times (~0,~1)\times (~1,~2)$ & (~1,~1,~1) & (~1,~1,~0)\\
    
     1 & b1 & $(~1,-2)\times (~0,-1)\times (-1,~1)$ & (~1,~1,~1)  & (~1,~1,~0)\\
     
     1 & b2 & $(~1,-2)\times (~0,-1)\times (-1,~1)$ & (~1,~1,~1)  & (~1,~1,~1)\\
     
     2 & c & $(-1,~2)\times (~0,-1)\times  (~1,-1)$ & (~1,~1,~1)  & (~1,~1,~1)\\
     
\hline
\hline

     9 & d1 & $(~1,~1)\times (~0,-1)\times (~1,~2)$ & (~1,~1,~1) & (~1,~1,~1)\\
     
     3 & d2 & $(~1,~1)\times (~0,-1)\times (~1,~2)$ & (~1,~1,~1) & (~1,~1,~0)\\
    
     3 & d3 & $(~1,~2)\times (~0,~1)\times (-1,-1)$ & (~1,~1,~1)  & (~1,~1,~1)\\
     
     3 & d4 & $(~1,~2)\times (~0,~1)\times (-1,-1)$ & (~1,~1,~1)  & (~1,~1,~0)\\
     
\hline
\hline     
     
     3 & e1 & $(~1,~0)\times (~0,-1)\times  (~0,~1)$ & (~1,~1,-1)  & (~1,~1,~1)\\
     
     1 & e2 & $(~1,~0)\times (~0,-1)\times  (~0,~1)$ & (~1,~1,-1)  & (~1,~1,~0)\\
     
     7 & e3 & $(-1,~0)\times (~0,~1)\times  (~0,~1)$ & (~1,~1,~1)  & (~1,~1,~1)\\
     
     7 & e4 & $(-1,~0)\times (~0,~1)\times  (~0,~1)$ & (~1,~1,~1)  & (~1,~1,~0)\\
     
     5 & e5 & $(~0,~1)\times (~0,~1)\times  (-1,~0)$ & (~1,~1,~1)  & (~1,~1,~1)\\
     
     11 & e6 & $(~0,~1)\times (~0,~1)\times  (-1,~0)$ & (~1,-1,~1)  & (~1,~1,~1)\\ 
     
     5 & e7 & $(~0,~1)\times (~0,~1)\times  (-1,~0)$ & (~1,~1,~1)  & (~0,~1,~1)\\
     
     5 & e8 & $(~0,-1)\times (~0,~1)\times  (~1,~0)$ & (~1,-1,~1)  & (~0,~1,~1)\\

\hline
\hline
  
      4 & $\beta_1$ & $(~1,-1)\times (~1,~0)\times (~1,~1)$ & \mbox{bulk} & \mbox{bulk} \\
      
      6 & $\beta_2$ & $(~1,~0)\times (~2,-1)\times (~1,~1)$ & \mbox{bulk} & \mbox{bulk} \\
      
      96 & 1 & $(~1,~0)\times (~1,~0)\times (~1,~0)$ & \mbox{bulk} & \mbox{bulk} \\
     
     14 & 1 & $(~1,~0)\times (~0,-1)\times (~1,~1)$ & \mbox{bulk} & \mbox{bulk} \\
     
     76 & 2 & $(~0,-1)\times (~1,~0)\times (~0,~1)$ & \mbox{bulk} & \mbox{bulk} \\
     
     24 & 3 & $(~0,-1)\times (~0,~1)\times (~1,~0)$ & \mbox{bulk}&  \mbox{bulk} \\
\hline          

\end{tabular}
\end{center}
\end{table}

\begin{table}[htb]
\footnotesize
\renewcommand{\arraystretch}{1.0}
\caption{Intersection numbers for the MSSM sector of the model with the D6-brane configurations shown in Table~\ref{rigidmodel3gen_MSSM}.}
\label{SMintnum}
\begin{center}
\begin{tabular}{|c||c|c||c|c|c|c|c|c|c|c|c|c|c|c|}
\hline
& \multicolumn{11}{c|}{$SU(3)_C\times SU(2)_L\times U(1)_Y$}\\
\hline \hline  & $N$ & $n_{S}$& $n_{A}$ & $a2$ & $a2'$ & $b1$ & $b1'$ & $b2$ & $b2'$ & $c$ & $c'$ \\

\hline

    $a1$&  3&  0 &  6  & 0 & -4 & 3 & 0 & 3 & 0 & -3 & 0 \\
    
    $a2$&  1&  0 &  6  & - & - & 3 & 0 & 3 & 0 & -3 & 0 \\

    $b1$&  1&  0 & -6  & - & - & - & - & 2 & 0 & 0 & -6 \\
    
    $b2$&  1&  0 & -6  & - & - & - & - & - & - & 0 & -2 \\

    $c$&  2&  0 & -6  & - & - & - & - & - & - & - & -  \\

\hline

\end{tabular}

\end{center}

\end{table}

The gauge bosons for almost all $U(1)$ groups will become massive as can be
seen from the GS cancellation conditions Eqs.~(\ref{twistedGS}) and (\ref{GSeq}).
After 
splitting the stacks, the $U(1)$ factors
\begin{eqnarray}
U(1)_{B-L} = \frac{1}{6}(U(1)_{a1} - 3U(1)_{a2}), \\ \nonumber
U(1)_{I3R} = \frac{1}{2}(U(1)_{b1} - U(1)_{b2}),
\end{eqnarray}
survive the GS conditions Eq.~(\ref{GSeq}).  If the relevant D6-branes were wrapping bulk cycles, this would
gaurantee that the gauge bosons associated with these groups
would remain massless.  However, since stack $a1$ and $a2$ do not pass through
the same fixed points on the third torus, $U(1)_{B-L}$ is anomalous due to the additional constraints Eq.~(\ref{twistedGS}). Similar considerations apply to $U(1)_{I3R}$.  For phenomenological reasons, this might actually desirable.  In particular, it is not possible
to generate couplings such as a Majorana neutrino mass term
$W_M = \lambda_i N N$
via D-brane instantons if $U(1)_{B-L}$ remains gauged at the string scale.   

\begin{table}[tb]

\footnotesize

\renewcommand{\arraystretch}{1.0}

\caption{The chiral and vector-like superfields for states charged MSSM gauge groups, and their quantum
numbers under the gauge symmetry $SU(3)_C\times 
SU(2)_L \times U(1)_Y \times SU(4)^4 \times SU(6) \times \prod_{i=1}^3 Usp(N_i)$.}

\label{MSSMspec}

\begin{center}

\begin{tabular}{|c||c||c|c|c|c|c|c|c||c|c|}\hline

 & Quantum Number

& $Q_{a1}$ & $Q_{a2}$  & $Q_{b1}$ & $O_{b2}$ & $Q_{c}$ & $Q_Y$ & Field \\

\hline\hline

$a1b1$ & $3\times (\overline{3},1,1,1,1,1,1,1,1,1,1,1,1)$ & -1 & 0 & 1 & 0 & 0  & 1/3  & $D^c$\\

$a1b2$ & $3\times (\overline{3},1,1,1,1,1,1,1,1,1,1,1,1)$ & -1 & 0 & 0 & 1 & 0  &-2/3  & $U^c$\\

$a1c$ & $3 \times (3,1,\overline{2},1,1,1,1,1,1,1,1,1,1)$ & 1 & 0  & 0 & 0 & -1 & 1/6  & $Q_L$\\

$a2b1$ & $3\times (1,1,1,1,1,1,1,1,1,1,1,1,1)$            & 0 &-1  & 1 & 0 & 0  &  1   & $E^c$\\

$a2b2$ & $3\times (1,1,1,1,1,1,1,1,1,1,1,1,1)$ & 0 &-1  & 0 & 1 & 0  &  0   & $N$\\

$a2c$ & $3 \times (1,1,1,1,\overline{2},1,1,1,1,1,1,1,1)$ & 0 & 1  & 0 & 0 & -1 &-1/2  & $L$\\

$b1c$ & $6 \times (1,1,1,1,2,1,1,1,1,1,1,1,1)$ & 0 & 0  & -1 & 0 & 1 &-1/2  & $H_d^i$\\

& $6 \times (1,1,1,1,\overline{2},1,1,1,1,1,1,1,1)$ & 0 & 0 & 1  & 0  & -1 & 1/2  & $\overline{H}_d^i$ \\ 

$b2c$ & $6 \times (1,\overline{2},2,1,1,1,1,1,1,1,1,1,1)$ & 0 & 0  & 0 & -1 & 1 & 1/2  & $H_u^i$\\

& $6 \times (1,2,\overline{2},1,1,1,1,1,1,1,1,1,1)$ & 0 &  0 & 0  & 1 & -1 & -1/2  & $\overline{H}_u^i$ \\ 

$b1'b2$ & $2 \times (1,1,1,1,1,1,1,1,1,1,1,1,1)$ & 0 & 0  & -1 & -1 & 0 & 0  & $S$\\
\hline

$b1'c$ & $6 \times (1,1,1,1,2,1,1,1,1,1,1,1,1)$ & 0 & 0 & 1  & 0  & 1 & 1/2 & $H_1^i$\\

$b2'c$ & $2 \times (1,1,1,1,2,1,1,1,1,1,1,1,1)$ & 0 & 0 & 0  & 1  & 1 & -1/2 & $H_1^i$\\

\hline
\end{tabular}

\end{center}

\end{table}

Although it may be desirable for $U(1)_{B-L}$ to become massive, 
care must be taken to ensure that the SM hypercharge, given by 
\begin{equation}
U(1)_{Y_0} = \frac{1}{6}(U(1)_{a1} - 3U(1)_{a2} + 3U(1)_{b1} - 3U(1)_{b2})
\label{hyper}
\end{equation}
does not also. Although this linear combination for the hypercharge will obviously satisfy Eq.~(\ref{GSeq}), one may easily see that $U(1)_Y$ will remain massless only for very special conditions due to the additional constraints.  Indeed, Eq.~(\ref{hyper}) does in fact become massive for the present model.  In general, the only chance for the combination given by Eq.~(\ref{hyper}) to remain massless is if the wrapping numbers and twisted charge assignments contrive
in such a way that the GS conditions can be satisfied.  Alternatively, the definition of the hypercharge may be extended to include
$U(1)$ groups from other stacks of branes.  Ideal candidates for such additional stacks are those wrapping bulk cycles which are invariant 
under the orientifold action since these will automatically satisfy Eq.~(\ref{GSeq}). Clearly the complete fractional cycles of such stacks
must not be invariant under the orientifold action so that a stack of $N$ of these D6-branes has a gauge group $U(N) = SU(N)\times U(1)$ in its worldvolume.  
After splitting the stacks SM hypercharge may remain massless provided that we redefine the SM hypercharge to be 
\begin{eqnarray}
U(1)_Y = \frac{1}{6}(U(1)_{a1} - 3U(1)_{a2} + 3U(1)_{b1} - 3U(1)_{b2}+ \\ \nonumber
\frac{1}{3}U(1)_{d1}-U(1)_{d2}-U(1)_{d3}+U(1)_{d4}).
\label{hyperp}
\end{eqnarray}
Obviously, this is not especially desirable for phenomenological reasons.  However, for the present we consider this as a possible
solution and defer further consideration of this issue.

\subsection{The Question of Gauge Coupling Unification}

The MSSM predicts the unification of the three gauge couplings at
an energy $\sim2.4\times10^{16}$~GeV. In intersecting D-brane
models, the gauge groups arise from different stacks of branes,
and so they will not generally have the same volume in the
compactified space. Thus, the gauge couplings are not
automatically unified, in contrast to heterotic models. 
For this reason, MSSM-like models where the gauge couplings 
happen to be unified appears to be a coincidence; there is 
no apparent deep reason for them to be unified is these types of
models, it just works out this way.  On the other hand, Pati-Slam models
where the D-branes of the observable sector all wrap bulk cycles which are homologically 
the same automatically results in tree-level gauge coupling unification at
the string scale.  For this class of models, the apparent unification seems less coincidental since
it is possible to understand from where it emerges.   

The holomorphic gauge kinetic
function for a D6-brane wrapping a calibrated three-cyce is given
by~\cite{Blumenhagen:2006ci}
\begin{equation}
f_P = \frac{1}{2\pi \ell_s^3}\left[e^{-\phi}\int_{\Pi_P} \mbox{Re}(e^{-i\theta_P}\Omega_3)-i\int_{\Pi_P}C_3\right].
\end{equation}
In terms of the three-cycle wrapped by the stack of branes, we have
\begin{equation}
\int_{\Pi_a}\Omega_3 = \frac{1}{4}\prod_{i=1}^3(n_a^iR_1^i + im_a^iR_2^i).
\end{equation}
from which it follows that
\begin{eqnarray}
f_P &=&
\frac{1}{4\kappa_P}(n_P^1\,n_P^2\,n_P^3\,s-n_P^1\,m_P^2\,m_P^3\,u^1 -n_P^2\,m_P^1\,m_P^3\,u^2-
n_P^3\,m_P^1\,m_P^2\,u^3),
\label{kingauagefun}
\end{eqnarray}
where $\kappa_P = 1$ for $SU(N_P)$ and $\kappa_P = 2$ for
$USp(2N_P)$ or $SO(2N_P)$ gauge groups and where we use the $s$ and
$u$ moduli in the supergravity basis.  In the string theory basis,
we have the dilaton $S$, three K\"ahler moduli $T^i$, and three
complex structure moduli $U^i$~\cite{Lust:2004cx}. These are related to the
corresponding moduli in the supergravity basis by
\begin{eqnarray}
\mathrm{Re}\,(s)& =&
\frac{e^{-{\phi}_4}}{2\pi}\,\left(\frac{\sqrt{\mathrm{Im}\,U^{1}\,
\mathrm{Im}\,U^{2}\,\mathrm{Im}\,U^3}}{|U^1U^2U^3|}\right)
\nonumber \\
\mathrm{Re}\,(u^j)& =&
\frac{e^{-{\phi}_4}}{2\pi}\left(\sqrt{\frac{\mathrm{Im}\,U^{j}}
{\mathrm{Im}\,U^{k}\,\mathrm{Im}\,U^l}}\right)\;
\left|\frac{U^k\,U^l}{U^j}\right| \qquad (j,k,l)=(\overline{1,2,3})
\nonumber \\
\mathrm{Re}(t^j)&=&\frac{i\alpha'}{T^j} \label{idb:eq:moduli}
\end{eqnarray}
and $\phi_4$ is the four-dimensional dilaton.

The gauge coupling constant associated with a stack P is given by
\begin{eqnarray}
g_{D6_P}^{-2} &=& |\mathrm{Re}\,(f_P)|.\label{idb:eq:gkf}
\end{eqnarray}
For the model under study the $SU(3)$ holomorphic
gauge function is identified with stack $a1$ and the $SU(2)$
holomorphic gauge function with stack $c$. The $U(1)_Y$ holomorphic
gauge function is then given by taking a linear combination of the
holomorphic gauge functions from all the stacks. If we consider the definition of the hypercharge
given by Eq.~(\ref{hyper}), then we find that each of the stacks in the observable sector turns out to have the same gauge coupling. 
Thus, the couplings will
be automatically unified at the string scale with canonical normalization on the hypercharge.  
\begin{equation}
g^2_{s} = g^2_{w} = \frac{5}{3}g^2_Y.
\label{gaugecouplings}
\end{equation}
However, we recall the the combination given in Eq.~(\ref{hyper}) becomes massive due to the twisted
Green-Schwarz conditions, and was
redefined in Eq.~(55) to include the additional stacks $d_i$, $i=1-4$, thus we then actually have  
\begin{equation}
g^2_{s} = g^2_{w} = (\frac{5}{3}+\Delta)g^2_Y,
\label{gaugecouplings2}
\end{equation}
where $\Delta$ is an effective threshold correction which expresses the effect of the stacks $d_i$ on the 
normalization of the hypercharge.  For the present model, we have $\Delta=5/9$ which disrupts the 
unification of $g^2_{s}$ and $g^2_{w}$ with $g^2_Y$.  However, it may be possible to find an alternative
combination of stacks which may be added to the hypercharge in such a way as 
to ameliorate this problem.  Since we do not require the actual unification, only the 
apparent unification, we only need $\Delta$ to be a small value. 
Although the model is a Pati-Salam \lq GUT\rq \ at the string scale, the gauge unification
need not occur in such models.  It is desirable only because this
seems to be what is observed;  gauge unification appears to be a coincidence.

\section{Conclusion}

In this paper, we have constructed three-family
Pati-Salam and MSSM-like models in AdS as Type IIA flux vacua on the $T^6/(\Z_2 \times \Z_2')$
orientifold.  For each of these models, the D-branes are wrapping rigid
cycles, which freezes the open-string moduli which correspond to the D-brane positions
and Wilson lines. In intersecting D-brane models where the D-branes wrap only bulk
cycles, there arises matter in the adjoint representation which results from 
unstabilized open-string moduli.  As a result, there are light-scalars present in
such models which are charged under the MSSM gauge groups.  Besides being phenomenologically
undesirable since such scalars are not experimentally observed, such fields
can have a negative effect on the running of the gauge couplings.  In particular, 
the asymptotic freedom of $SU(3)_C$ can be destroyed, as well as the asyptotic freedom
of hidden sector gauge groups.  Typically, there is exotic matter present in intersecting
D-brane models which are charged under both the MSSM and hidden sector gauge groups, and
one way of generating large masses for such states is if the hidden sectors groups become confining
at some high scale which requires that these groups have negative $\beta$ functions.  One of the effects of unstabilized open-string
moduli is  therefore to ruin this possibility.  

For the MSSM-like model we considered, we found that the tree-level gauge couplings associated with 
$SU(3)_C$ and $SU(2)_L$ may be unified at the string scale.  However, we found that in order for 
the hypercharge $U(1)_Y$ to remain massless, it is necessary to extend to the definition of the
hypercharge to include $U(1)$ factors from other stacks of D6-branes.  Besides being undesirable
since this increases the liklihood that there will be extra matter charged under all three MSSM
gauge group, it also has the effect of shifting the hypercharge away from a canonical normalization
in order to be unified with the other two gauge groups.  This shift, which is like an effective
threshold correction, depends on the details of what is added to the hypercharge in order to keep
it massless, in which there is some freedom.  

Besides the reasons given above which motivate constructing models with frozen open-string moduli,
it is also the case that the Yukawa couplings which must be present in the superpotential to generate masses
for the quarks and leptons depend directly on the open-string moduli.  Previously, for the specific
three-generation intersecting D-brane model discussed in~\cite{Chen:2007px} and~\cite{Chen:2007zu} it is possible to obtain correct masses and mixings
for both the up and down-type quarks, as well as the tau lepton considering only the trilinear couplings.  In addition,
it is possible to obtain the correct electron and muon masses, in general, by considering contributions to the Yukawa couplings from four-point functions~\cite{Chen:2008rx}.  Despite the impressive successes of this model, the open-string moduli
were not stabilized and so it was not possible to obtain unique calculations of the Yukawa coupligns.  Essentially,
the open-string moduli VEVs were treated as free parameters.   For the models we considered in this paper, the mass hierarchies can arise in principle from D-brane instanton induced Yukawa couplings.  Since all of the D6-branes and the E2-brane associated with the instanton are wrapping rigid
cycles, all of the open-string moduli are fixed and there is limited freedom to \lq tune\rq \ the couplings to
give the desired hierarchies. It would be very interesting to see if the observed mass heirarchies for quarks 
and leptons can be obtained for this model.  We plan to explore this fully in future work.

\newpage
\section{Acknowledgements}
The work of C.M. is supported by the Mitchell-Heep Chair in High Energy Physics.  The work of D.V. Nanopoulos and V.E. Mayes is supported by DOE grant DE-FG03-95-Er-40917.  The work of T. Li was supported in part by the Cambridge-Mitchell Collaboration in Theoretical Cosmology and by the Natural Science Foundation of China under grant number 10821504.

\end{document}